\documentclass[11pt]{article}
\usepackage[OT1]{fontenc}
\usepackage[utf8]{inputenc}
\usepackage{graphicx}
\usepackage[margin=1in]{geometry}
\usepackage[bf,small,raggedright,compact]{titlesec}
\usepackage{paralist}
\usepackage[longnamesfirst,numbers,sort&compress]{natbib}
\usepackage{stmaryrd}

\usepackage{amsmath,amssymb,amsthm}
\usepackage{tikz}
\usepackage{algorithm, algpseudocode}
\usepackage{enumitem}
\usepackage{xcolor}
\usepackage[font=small,labelfont=bf,labelsep=period,skip=6pt]{caption}
\usepackage{thmtools}
\makeatletter
\@ifundefined{newcounteralias}{}{%
  \renewcommand\thmt@autorefsetup{%
    \@xa\def\csname\thmt@envname autorefname\@xa\endcsname\@xa{\thmt@thmname}%
  }%
}
\makeatother
\usepackage{hyperref}

\definecolor{mipInk}{HTML}{263745}
\definecolor{mipGray}{HTML}{586871}
\definecolor{mipBlue}{HTML}{146480}
\definecolor{mipRust}{HTML}{A54823}
\definecolor{mipBase}{HTML}{E9EDF0}
\definecolor{mipPaleBlue}{HTML}{CBE4EB}
\tikzset{
  mip picture/.style={font=\small,line cap=round,line join=round,inner sep=2pt,text=mipInk},
  mip panel/.style={font=\small\bfseries,inner sep=1pt},
  mip boundary/.style={draw=mipInk,line width=0.9pt},
  mip input/.style={draw=mipGray,fill=mipGray!22,line width=0.7pt},
  mip pyramid/.style={draw=mipBlue,fill=mipPaleBlue,line width=1.0pt},
  mip chain/.style={draw=mipRust,line width=1.4pt},
  mip bend/.style={rectangle,fill=mipRust,inner sep=2.0pt},
  mip leader/.style={draw=mipGray,line width=0.55pt}
}

\theoremstyle{plain}
\newtheorem{theorem}{Theorem}
\newtheorem{lemma}[theorem]{Lemma}

\theoremstyle{definition}

\theoremstyle{remark}

\newcommand{\Vol}{\mathrm{Vol}}
\newcommand{\Surf}{\mathrm{Surf}}
 
\DeclareMathOperator{\rank}{rank}

\newcommand{\R}{\mathbb{R}}
\newcommand{\B}{\mathsf{B}}
\newcommand{\cS}{\mathcal{S}}

\newcommand{\conv}{\operatorname{conv}}
\newcommand{\cof}{\operatorname{cof}}
\newcommand{\surf}{\operatorname{surf}}
\newcommand{\vol}{\operatorname{vol}}
\newcommand{\OPT}{\mathrm{OPT}}
\newcommand{\poly}{\operatorname{poly}}
\newcommand{\eps}{\varepsilon}

\title{Approximating a Minimum Intersecting Convex Polytope}
\author{Sándor Kisfaludi-Bak\thanks{Department of Computer Science, Aalto University, Espoo, Finland. Emails: sandor.kisfaludi-bak@aalto.fi and
Saeed.Odak@aalto.fi. Supported by the Research Council of Finland, Grant 363444.}, \quad Beimar J. Naranjo\thanks{
Eötvös Loránd University, Budapest, Hungary. Email: beimar@student.elte.hu. This work was carried out while this author was an AScI summer research intern at Aalto University, Espoo, Finland, 2026.
}, \quad Saeed Odak\footnotemark[1]}
\date{\today}
\begin{document}
\maketitle
\begin{abstract}
In fixed dimension, we give efficient polynomial-time approximation schemes for minimizing the volume or surface area of a convex polytope that intersects every member of a given family of closed semialgebraic sets of bounded degree.
Our algorithms combine a geometric coreset argument with the real algebraic optimization technique of Basu, Pollack, and Roy.
Assuming the Satisfiability Hypothesis we show that the dependence on the approximation parameter $\varepsilon$ is nearly optimal.
\end{abstract}
\section{Introduction}
\label{sec:introduction}

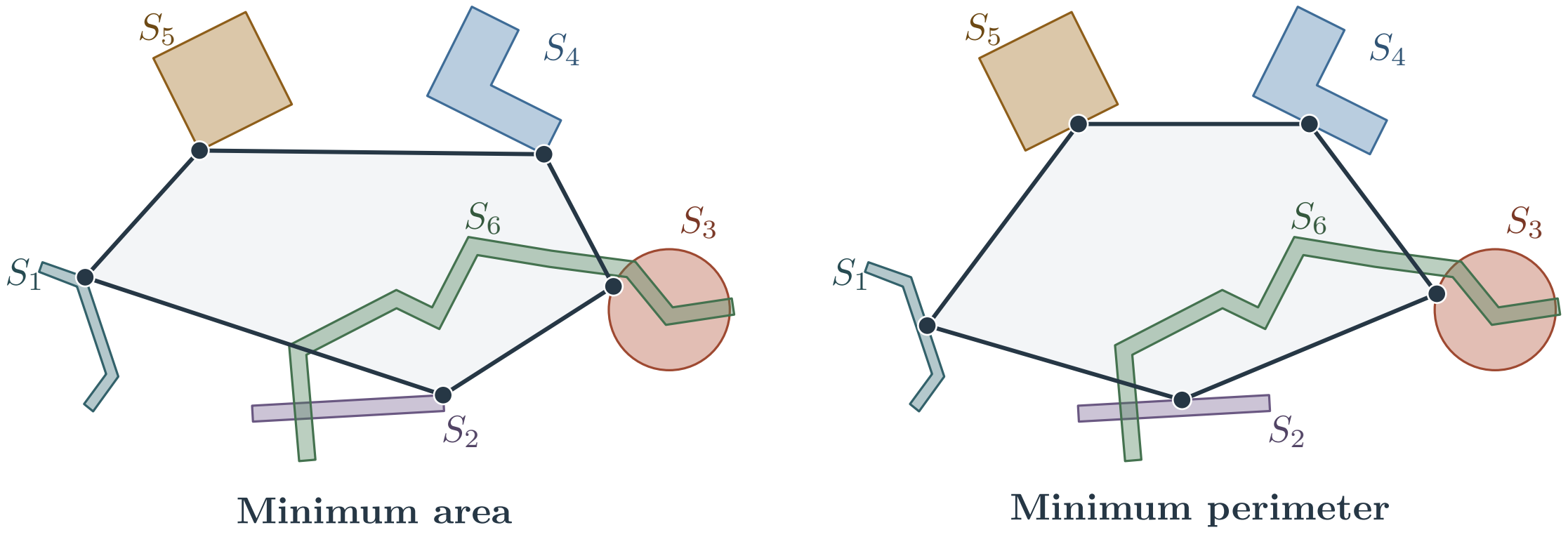
\begin{figure}[!htbp]
\centering
\begin{tikzpicture}[mip picture,x=1cm,y=1cm,
  region/.style={draw=#1!88!black,fill=#1,fill opacity=0.38,line width=0.6pt},
  hull/.style={draw=mipInk,line width=1.12pt},
  contact/.style={circle,draw=white,line width=0.5pt,fill=mipInk,inner sep=1.75pt}]
  \definecolor{regionTeal}{HTML}{396E78}
  \definecolor{regionViolet}{HTML}{786494}
  \definecolor{regionRust}{HTML}{B45338}
  \definecolor{regionBlue}{HTML}{477BAC}
  \definecolor{regionGold}{HTML}{A16B1D}
  \definecolor{regionGreen}{HTML}{4D825A}
  \def\inputregions{%
    \path[region=regionTeal] ({-4507/1500},{671/1500})--(-2.602,0.306)--(-2.29,-0.63)--({-3791/1500},-0.958)--
      ({-3919/1500},{-1333/1500})--({-3607/1500},-0.606)--({-4019/1500},0.218)--({-911/300},{107/300})--
      cycle;
    \path[region=regionViolet] (-1.03,{-407/450})--(0.77,{-181/225})--({467/600},{-859/900})--({-613/600},{-949/900})--
      cycle;
    \path[region=regionRust] (2.9,0) circle[radius={sqrt(13/40)}];
    \path[region=regionBlue] (1.72,1.465)--(0.62,2.015)--(1.04,2.855)--(1.48,2.635)--
      (1.22,2.115)--(1.88,1.785)--cycle;
    \path[region=regionGold] (-0.655,1.935)--(-1.525,1.5)--(-1.96,2.37)--(-1.09,2.805)--
      cycle;
    \path[region=regionGreen] (-0.5867,-1.4266)--(-0.6812,-0.3352)--(0.329,0.1861)--(0.666,0.0222)--
      (1.0072,0.6851)--(1.7913,0.5562)--(2.5803,0.4524)--(2.9319,0.0226)--
      (3.4886,0.1061)--(3.5114,-0.0461)--(2.8681,-0.1426)--(2.4997,0.3076)--
      (1.7687,0.4038)--(1.0928,0.5149)--(0.734,-0.1822)--(0.331,0.0139)--
      (-0.5188,-0.4248)--(-0.4333,-1.4134)--cycle;
  }
  \def\regionlabels{%
    \node[text=regionTeal!68!black] at (-3.17,0.33) {$S_1$};
    \node[text=regionViolet!68!black] at (0.94,-1.15) {$S_2$};
    \node[text=regionRust!68!black] at (3.18,0.82) {$S_3$};
    \node[text=regionBlue!68!black] at (1.89,2.45) {$S_4$};
    \node[text=regionGold!68!black] at (-1.92,2.64) {$S_5$};
    \node[text=regionGreen!68!black] at (1.15,0.86) {$S_6$};
  }
  \def\pentagon{(p1)--(p2)--(p3)--(p4)--(p5)--cycle}
  \def\drawinstance{%
    \fill[mipBase!55] \pentagon;
    \inputregions
    \draw[hull] \pentagon;
    \foreach \v in {p1,p2,p3,p4,p5} \node[contact] at (\v) {};
    \regionlabels
  }
  \begin{scope}
    \coordinate (p1) at (-2.602,0.306);
    \coordinate (p2) at (0.77,{-181/225});
    \pgfmathsetmacro{\diskstep}{sqrt((13/40)/((817/360)^2+(19/20)^2))}
    \coordinate (p3) at ({2.9-\diskstep*(817/360)},{\diskstep*(19/20)});
    \coordinate (p4) at (1.72,1.465);
    \coordinate (p5) at (-1.525,1.5);
    \drawinstance
    \node[mip panel] at (0.12,-1.89) {Minimum area};
  \end{scope}
  \begin{scope}[xshift=7.78cm]
    \coordinate (p1) at (-2.45,-0.15);
    \coordinate (p2) at (-0.05,-0.85);
    \coordinate (p3) at (2.35,0.15);
    \coordinate (p4) at (1.15,1.75);
    \coordinate (p5) at (-1.025,1.75);
    \drawinstance
    \node[mip panel] at (0.12,-1.89) {Minimum perimeter};
  \end{scope}
\end{tikzpicture}%
%
\caption{Optimal intersecting pentagons for the same six regions in the place.}
\label{fig:overlapping-instance}
\end{figure}

Computing the convex hull of a finite point set is one of the central problems that shaped computational geometry. In every fixed dimension it is well understood: optimal algorithms have long been known in the plane and in three dimensions~\cite{Graham1972,PreparataHong1977}, and Chazelle~\cite{Chazelle93} gave a worst-case optimal algorithm in every fixed dimension. The convex hull is the smallest convex set containing the points. Since volume and surface area are monotone under inclusion for convex sets, it minimizes both measures at the same time among all convex sets that contain the points. In many applications, however, the points are not known exactly. Coordinates may come from measurements with bounded error or may be stored with limited precision. This is the model of \emph{imprecise points}~\cite{Loffler2009thesis,LofflerVanKreveld2010}, in which each point is specified by a region of possible locations and one asks for the best or worst value of a geometric measure over all realizations. For the convex hull, the most basic such question asks to select one location from each region such that the convex hull of the selected points has minimum volume or minimum surface area. Equivalently, we seek a convex polytope of minimum volume or surface area that intersects every region. In contrast to the precise setting, the two objectives generally have different optimal solutions, so they lead to two different problems~\cite{ABKS25}. Figure~\ref{fig:overlapping-instance} shows a planar instance whose minimum-area and minimum-perimeter solutions are distinct pentagons. Four contacts move from corners of the input regions to the interiors of edges when the objective changes from area to perimeter.

In the plane, volume and surface area become area and perimeter, and the problems have been studied both for imprecise points and as polygon intersecting problems. Rappaport~\cite{Rappaport1995} introduced the minimum-perimeter intersecting polygon problem for segments. He gave an $O(n\log n)$-time exact algorithm when the segments have a constant number of orientations and an exponential-time algorithm in general, and Hassanzadeh and Rappaport~\cite{HassanzadehRappaport2009} later obtained a polynomial-time $\pi/2$-approximation. L\"offler and van Kreveld~\cite{LofflerVanKreveld2010} systematically studied smallest and largest convex hulls of imprecise points modeled as squares, segments, and disks. For axis-parallel squares, they gave exact algorithms for minimum area and minimum perimeter running in $O(n^2)$ and $O(n\log n)$ time, respectively, and they proved NP-hardness for some maximization variants; approximation schemes for these were obtained with coresets~\cite{LofflerVanKreveld2008}. For arbitrarily oriented segments, Dumitrescu and Jiang~\cite{DumitrescuJiang2012} gave a $4(1+\eta)/\pi$-approximation for every $\eta>0$, including a $1.28$-approximation from a constant number of linear programs. They also gave a polynomial-time approximation scheme (PTAS) for minimum perimeter, based on discretization and enumeration, with running time $n^{O(1)}/\varepsilon+2^{O(\varepsilon^{-2/3})}n$. In the other direction, they proved that the problem is NP-hard for nonconvex simple polygons and for polygonal chains, by a reduction from \textsc{Vertex Cover}.

Antoniadis, de Berg, Kisfaludi-Bak, and Skarlatos~\cite{ABKS25} later gave fully polynomial-time approximation schemes (FPTASes) for minimum area and minimum perimeter on possibly intersecting convex polygons. Their algorithms show that a near-optimal solution with vertices on a suitably placed grid exists, and they find one by dynamic programming along the boundary of the solution, using orderings of the tangents to the convex input polygons to decide which objects each subproblem must meet. They also gave an exact polynomial-time algorithm for minimum perimeter on possibly intersecting segments. Both the dynamic programs and the tangent orderings rely on the planar structure of the problem. Antoniadis et al.\ asked whether minimum-volume and minimum-surface-area intersecting polytopes of convex polytopes in $\R^3$ admit a PTAS.

Geometric optimization problems in which a solution must only meet every input region, rather than pass through prescribed points, have been studied extensively in the literature. In the traveling salesman problem with neighborhoods (TSPN), introduced by Arkin and Hassin~\cite{ArkinHassin1994}, a shortest tour must visit every input region. De Berg et al.~\cite{deBergEtAl2005} gave a constant-factor approximation for disjoint fat convex regions of arbitrary size, and showed that TSPN is APX-hard when the regions may overlap and need not be convex or fat. When the regions are lines in the plane, an optimal tour is convex, so TSPN coincides with the minimum-perimeter intersecting polygon problem for lines~\cite{DumitrescuJiang2012,ABKS25}. In higher dimensions, TSPN with hyperplanes admits a PTAS~\cite{AntoniadisFHS2020}, whereas TSPN with lines in $\R^d$ is APX-hard for every $d\geq3$~\cite{AntoniadisKisfaludiBakLaekhanukitVaz2022}. Similarly, for a connected collection of lines or line segments, the watchman route problem asks for a shortest closed curve contained in their union that intersects every input line or segment; it is polynomial-time solvable for lines in the plane, but NP-hard for lines in three dimensions and for segments in the plane~\cite{DumitrescuMitchellZylinski2014}. In the touring polygons problem, the regions must be visited in a prescribed order; it is polynomial-time solvable for convex polygons and NP-hard for nonconvex ones~\cite{DrorEtAl2003}, and it was recently revisited in the orthogonal setting~\cite{CaselEtAl2026}. All of these problems minimize the length of a route that meets the input objects, whereas we minimize the volume or surface area of a convex polytope that meets them. The convexity of the solution is what our algorithms exploit: it allows us to approximate an optimal solution by a polytope with few vertices, independently of the input. 

All approximation schemes above assume convex input sets, and their analyses use convexity in an essential way. Nonconvex inputs arise naturally. An imprecise point may be restricted to one of several disjoint regions, for instance when a measurement admits several consistent interpretations, and the union of these regions is nonconvex. Error bounds of measurements are also often curved regions, such as disks or ellipses, rather than polygons. For nonconvex polygons and polygonal chains the problem is already NP-hard in the plane~\cite{DumitrescuJiang2012}, so approximation is the natural goal. Dumitrescu and Jiang asked the following question~\cite{DumitrescuJiang2012}:

\begin{center}
\textbf{Question.} \emph{Does the minimum-perimeter intersecting polygon problem for simple polygons admit a polynomial-time approximation scheme, or at least a constant-factor approximation?}
\end{center}

We answer this question affirmatively in a strong form: we give efficient polynomial-time approximation schemes (EPTASes) for a much broader class of possibly nonconvex and possibly intersecting input sets, which includes simple polygons, and our algorithms work in every fixed dimension. Since convex polytopes in $\R^3$ belong to this class, this also gives a positive answer to the question of Antoniadis, de Berg, Kisfaludi-Bak, and Skarlatos~\cite{ABKS25}. We complement our algorithms with a conditional lower bound which shows that their dependence on $\varepsilon$ is nearly optimal.

\subsection{Our results}

Let $\mathcal S$ be a finite family of nonempty, closed semialgebraic sets in $\mathbb R^d$, each described by a quantifier-free Boolean formula involving polynomial equalities and inequalities of degree at most $r$. This class includes, among other examples, polytopes, balls, ellipsoids, polygonal chains, and finite unions of such sets. We denote by $n$ the total size of the input descriptions, counting one unit for each set and each predicate. Thus, if $S_i$ is described using $s_i$ predicates, then $|\mathcal S|+\sum_i s_i\leq n$. Throughout, we regard $d$ and $r$ as absolute constants. Although we optimize over all convex sets intersecting every member of $\mathcal S$, for each of the volume and surface-area objectives there is always an optimal solution that is a polytope. Thus, restricting attention to polytopes does not change either optimum. We give real-RAM approximation algorithms for both objectives.

\begin{theorem}[Approximation schemes]
\label{thm:intro-upper}
For fixed integers $d\geq2$ and $r\geq1$, \textnormal{\textsc{Minimum Volume Intersecting Polytope}} and \textnormal{\textsc{Minimum Surface-Area Intersecting Polytope}} admit deterministic $(1+\varepsilon)$-approximation algorithms on $\mathcal S$ with running time $2^{O_d(1/\varepsilon^{(d-1)/2}\log(1/\eps))}(nr)^{O(d^6)}$.
\end{theorem}

Our lower bounds are conditional on the Satisfiability Hypothesis (SH) of Stearns and Hunt~\cite{StearnsHunt1990}. For every fixed $0<\delta<1$, SH excludes a deterministic $2^{O(s^{1-\delta})}$-time algorithm for CNF satisfiability on inputs of size $s$. We measure the conditional lower bound in the standard bit model. Its starting point is the gap for Max-3SAT in~\cite[Corollary~5.1]{SHhardness}.

\begin{restatable}[Conditional lower bound]{theorem}{conditionalLowerBound}
\label{thm:intro-lower}\label{thm:lower-bound}
Assuming SH, for every fixed integer $d\geq2$ and every fixed $0<\delta<1$, neither \textnormal{\textsc{Minimum Volume Intersecting Polytope}} nor \textnormal{\textsc{Minimum Surface-Area Intersecting Polytope}} has a deterministic $(1+\varepsilon)$-approximation algorithm with running time $2^{O(1/\varepsilon^{(d-1-\delta)/2})}n^{O(1)}$ for all $0<\varepsilon<1$. This holds even when every input set is a point or a chain of two segments.
\end{restatable}

The running time contains a factor $2^{O_d(1/\varepsilon^{(d-1)/2}\log(1/\varepsilon))}$. Under SH, the lower bound rules out replacing $1/\varepsilon^{(d-1)/2}$ by any fixed smaller power of $1/\varepsilon$. In particular, it rules out a deterministic FPTAS for this input class. The hard instances use nonconvex chains, so the result is consistent with the FPTASes for convex polygons~\cite{ABKS25}.

\subsection{Overview of the techniques}

\begin{figure}[htbp]
\centering
\begin{tikzpicture}[mip picture,x=1cm,y=1cm,
  input size/.style={scale=1.20},
  region/.style={draw=#1!88!black,fill=#1,fill opacity=.38,line width=.6pt},
  hull/.style={draw=mipInk,line width=1.12pt},
  contact/.style={circle,draw=white,line width=.45pt,fill=mipInk,inner sep=1.62pt},
  bound/.style={draw=mipGray,densely dotted,line width=.78pt},
  object label/.style={font=\small,inner sep=1pt}]
  \definecolor{regionTeal}{HTML}{396E78}
  \definecolor{regionViolet}{HTML}{786494}
  \definecolor{regionRust}{HTML}{B45338}
  \definecolor{regionBlue}{HTML}{477BAC}
  \definecolor{regionGold}{HTML}{A16B1D}
  \def\leftTip{(-1.486719,-.105056)}
  \def\bottomTip{(-.206719,-1.615056)}
  \def\diskTip{(.807281,-.705056)}
  \def\rightTip{(1.573281,.424944)}
  \def\topTip{(-.166719,1.504944)}
  \def\optimum{\leftTip--\bottomTip--\diskTip--\rightTip--\topTip--cycle}
  \def\inscribed{\leftTip--\bottomTip--\rightTip--\topTip--cycle}
  \def\rounded{(-1.5,{-1/6})--({-1/6},{-10/6})--(1.5,.5)--({-1/6},1.5)--cycle}
  \def\inputsets{%
    \begin{scope}[shift/.expanded={\leftTip},input size]
      \path[region=regionTeal] (0,0)--(-.22,.20)--(-.72,.15)--(-.77,-.36)--
        (-.57,-.36)--(-.52,-.08)--(-.23,-.05)--cycle;
    \end{scope}
    \begin{scope}[shift/.expanded={\topTip},input size]
      \draw[regionViolet!88!black,line width=1.15pt] (-.50,.47)--(0,0)--(.54,.41);
    \end{scope}
    \begin{scope}[shift/.expanded={\rightTip},input size]
      \path[region=regionRust] (0,0)--(.21,-.04)--(.43,.22)--(.55,.19)--
        (.61,.40)--(.36,.48)--cycle;
    \end{scope}
    \begin{scope}[shift/.expanded={\bottomTip},input size]
      \path[region=regionBlue] (0,0)--(.38,-.16)--(.18,-.43)--
        (-.28,-.38)--(-.33,-.20)--cycle;
    \end{scope}
    \begin{scope}[shift/.expanded={\diskTip},input size]
      \path[region=regionGold] (.177070,-.154503) circle[radius=.235cm];
    \end{scope}
  }
  \def\setlabels{%
    \node[object label,text=regionTeal!68!black] at (-2.27,-.89) {$S_1$};
    \node[object label,text=regionViolet!68!black] at (.72,2.22) {$S_2$};
    \node[object label,text=regionRust!68!black] at (2.10,1.36) {$S_3$};
    \node[object label,text=regionBlue!68!black] at (-.98,-2.24) {$S_4$};
    \node[object label,text=regionGold!68!black] at (1.45,-1.24) {$S_5$};
  }
  \begin{scope}
    \fill[mipBase!55] \optimum;
    \inputsets
    \draw[bound] (0,0) circle[radius=1cm];
    \draw[hull] \optimum;
    \foreach \v in {\leftTip,\bottomTip,\diskTip,\rightTip,\topTip}
      \node[contact] at \v {};
    \node at (0,.1) {$K$};
    \setlabels
    \node[mip panel] at (0,-2.88) {(a) Normalized optimum};
  \end{scope}
  \begin{scope}[xshift=5.25cm]
    \fill[mipBase] \optimum;
    \fill[mipBase!42] \inscribed;
    \inputsets
    \draw[mipGray!80,line width=.65pt] \bottomTip--\diskTip--\rightTip;
    \begin{scope}[scale=0.83333333]
      \draw[bound] \optimum;
    \end{scope}
    \draw[hull] \inscribed;
    \foreach \v in {\leftTip,\bottomTip,\rightTip,\topTip}
      \node[contact] at \v {};
    \node at (0,.1) {$P$};
    \setlabels
    \node[mip panel] at (0,-2.88) {(b) Inscribed polygon};
  \end{scope}
  \begin{scope}[xshift=10.50cm,object label/.append style={fill=white}]
    \fill[mipBase!50] \optimum;
    \foreach \ix in {-14,...,14}
      \foreach \iy in {-15,...,14}
        \fill[mipGray!48] ({\ix/6},{\iy/6}) circle[radius=.31pt];
    \draw[mipGray!67,line width=.6pt] \optimum;
    \inputsets
    \draw[mipInk!85,dash pattern=on 3pt off 1.8pt,line width=.8pt] \rounded;
    \foreach \v in {(-1.5,{-1/6}),({-1/6},{-10/6}),(1.5,.5),({-1/6},1.5)}
      \node[rectangle,fill=mipInk,draw=white,line width=.36pt,inner sep=1.45pt] at \v {};
    \begin{scope}[scale=1.5]
      \draw[hull] \rounded;
    \end{scope}
    \node at (0,.1) {$Q$};
    \setlabels
    \node[mip panel] at (0,-2.88) {(c) Expanded grid polygon};
  \end{scope}
\end{tikzpicture}%
%
\caption{Approximating $K$ in normalized coordinates. (a)~$K$ contains the dotted unit disk and touches each of the five input sets at a marked vertex. (b)~The inscribed polygon $P$ contains the dotted copy $(1-\delta)K$ but misses the disk $S_5$. (c)~The square markers are the rounded vertices of $P$. They define the dashed grid polygon $Q$, whose expansion $(1-2\delta)^{-1}Q$ (solid outline) contains $K$ and meets all five sets.}
\label{fig:upper-overview}
\end{figure}
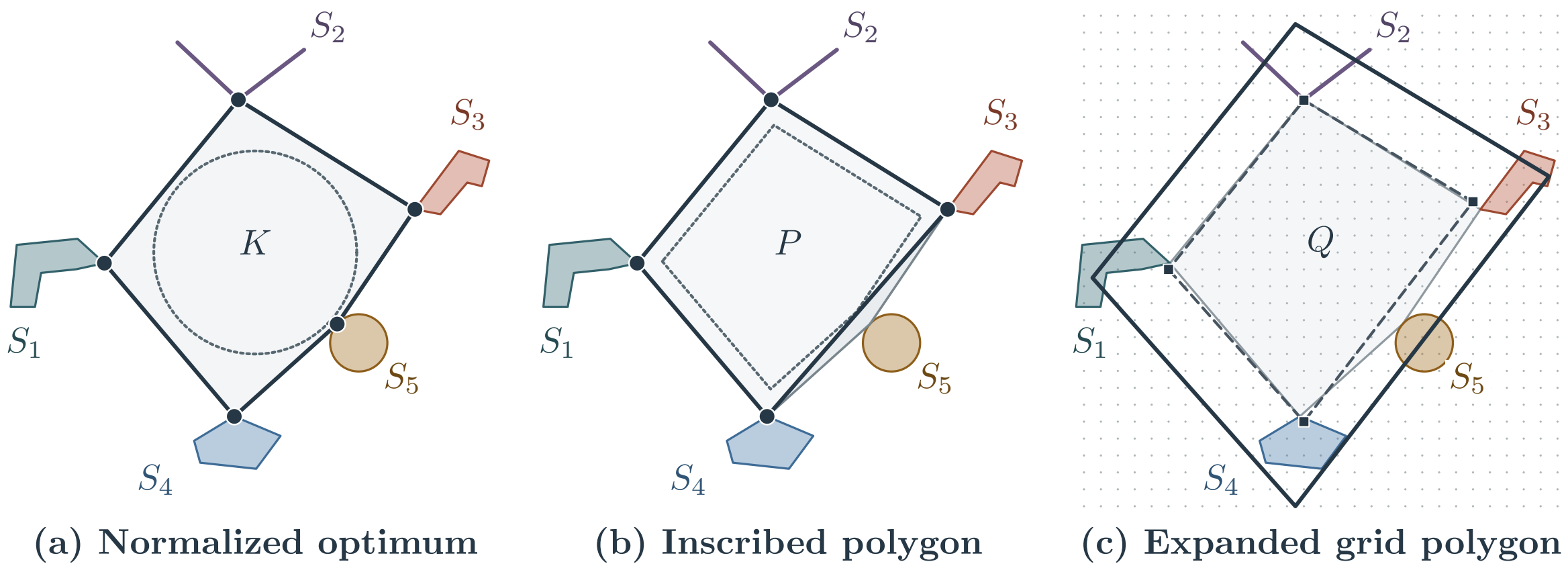

\paragraph{The approximation algorithms.}
We fix the dimension $d$ and consider a family $\cS$ of semialgebraic sets as described above. Our approach is essentially simple: first, we reduce the search to affine images of finitely many grid polytopes. Agarwal, Har-Peled, and Varadarajan~\cite{coreset} showed that a small coreset approximates the width of a finite point set in every direction. We use the containment formulation of this idea due to Nasz\'odi, Nazarov, and Ryabogin~\cite{NNR20}. Let $K$ be an optimal polytope with positive volume or respectively, positive surface area. We move its center of mass to the origin and choose an invertible linear map $T$ so that $T(K)$ contains the unit ball and lies in a ball whose radius depends only on $d$. Then, there is a polytope $P$ inscribed in $T(K)$ with $O_d((1/\varepsilon)^{(d-1)/2})$ vertices. It contains the copy of $T(K)$ scaled toward the origin by $1-O_d(\varepsilon)$. We round each vertex of $P$ to a grid point at distance at most $\varepsilon$ and call their convex hull $Q$. We show that scaling $Q$ outward from the origin by a factor of $1+O_d(\varepsilon)$ makes it contain $T(K)$. Applying $T^{-1}$ to this enlarged $Q$ gives a feasible affine image whose volume or surface area is at most $1+\varepsilon$ times the optimum. Figure~\ref{fig:upper-overview} illustrates why rounding and expansion together restore feasibility.

The grid has $O(\varepsilon^{-d})$ points. We enumerate its polytopes with $O(\varepsilon^{-(d-1)/2})$ vertices. For each grid polytope $Q$, we seek a feasible affine image with minimum volume or approximately minimum surface area. An image is feasible if it contains a point of each input set. We use quantifier elimination separately for each set to express this condition in terms of the affine map. The number of variables is therefore independent of the number of input sets. We then use the real algebraic optimization technique of Basu, Pollack, and Roy~\cite{BPR} to find the best image. Separating the intersection tests is what keeps this use of the general technique polynomial in the input size.

Volume is determined by the determinant of the affine map. The formula for surface area contains square roots. We approximate these square roots by a polynomial; a lower bound on the normalized surface-area sum controls the relative error. A separate feasibility test handles solutions of zero volume or zero surface area.

\paragraph{The lower bound.}
We reduce from \textsc{Vertex Cover} problem. Applying the standard reduction from satisfiability to independent set and assuming {Satisfiability Hypothesis (SH)} yields constants $0<\alpha<\beta<1$ such that no $2^{O(m^{1-\delta})}$-time algorithm distinguishes $m$-vertex graphs with a vertex cover of size at most $\alpha m$ from graphs in which every vertex cover has size at least $\beta m$. Our construction extends the planar idea of Dumitrescu and Jiang~\cite{DumitrescuJiang2012}, who placed the graph vertices at the vertices of a regular polygon. To obtain a lower bound that depends on the dimension, we must place many graph vertices, each contributing a small and equal amount to the cost, on the boundary of a polytope of constant size. Let $G$ be a graph with $m$ vertices, and set $L=2\lceil m^{1/(d-1)}\rceil$. Divide each coordinate of $[0,1]^{d-1}$ into $L$ equal intervals. Assign a different cell $C_v$ in $[0,1/2]^{d-1}$ to each vertex $v$ of $G$. Let $q$ be the piecewise-linear interpolation of the quarter-circle arc $\gamma(t)=1-\sqrt{1-t^2}$ at the interval endpoints, and put $\phi(x)=\sum_{i=1}^{d-1}q(x_i)$. For every corner $x$ of the cells, make $(x,\phi(x))$ a singleton input object. Their convex hull is the base polytope $K_0$, whose lower boundary has a facet $F_v$ above $C_v$, and every feasible polytope must contain $K_0$. At the center $c_v$ of that cell, place the apex $p_v=(c_v,\phi(c_v)-h_v)$, where $0<h_v\leq h:=1/(16L^2)$. The facet $F_v$ and apex $p_v$ bound a shallow pyramid. The increasing slopes of $q$ keep these pyramids separate, so their costs add when several apices are selected.

\begin{figure}[htbp]
\centering
\begin{tikzpicture}[mip picture,x=1.28cm,y=1.08cm]
  \coordinate (a) at (0,0);
  \coordinate (b) at (1.65,0.10);
  \coordinate (c) at (3.30,1.25);
  \coordinate (d) at (4.15,2.45);
  \coordinate (pu) at (0.825,-0.15);
  \coordinate (pv) at (2.475,0.475);
  \coordinate (w) at (3.30,-1.10);
  \begin{scope}
    \clip (-0.1,-0.25) rectangle (4.30,2.12);
    \fill[mipBase] (a)--(b)--(c)--(d)--(4.15,4.55)--cycle;
    \draw[mip boundary] (a)--(4.15,4.55) (a)--(b)--(c)--(d);
  \end{scope}
  \path[mip pyramid] (a)--(pu)--(b)--cycle;
  \path[mip pyramid] (b)--(pv)--(c)--cycle;
  \draw[mip chain] (pu)--(w)--(pv);
  \foreach \vertex in {a,b,c} \fill[mipInk] (\vertex) circle[radius=2.1pt];
  \foreach \vertex in {pu,pv} \fill[mipBlue] (\vertex) circle[radius=2.5pt];
  \node[mip bend] at (w) {};
  \node at (1.91,1.55) {$K_0$};
  \node[below left=2pt] at (pu) {$p_u$};
  \node[right=5pt,yshift=-1pt] at (pv) {$p_v$};
  \node[below right=1pt] at (w) {$w$};
  \node[mip panel] at (1.98,-1.82) {The edge object for $uv$};
\end{tikzpicture}%
%
\caption{The chain $[p_u,w]\cup[w,p_v]$ is the input object for the graph edge $uv$. The black singleton points force every feasible polytope to contain $K_0$. Meeting either arm of the chain then forces the polytope to contain the corresponding apex, $p_u$ or $p_v$. The pyramid depths are enlarged for visibility.}
\label{fig:lower-bound-overview}
\end{figure}
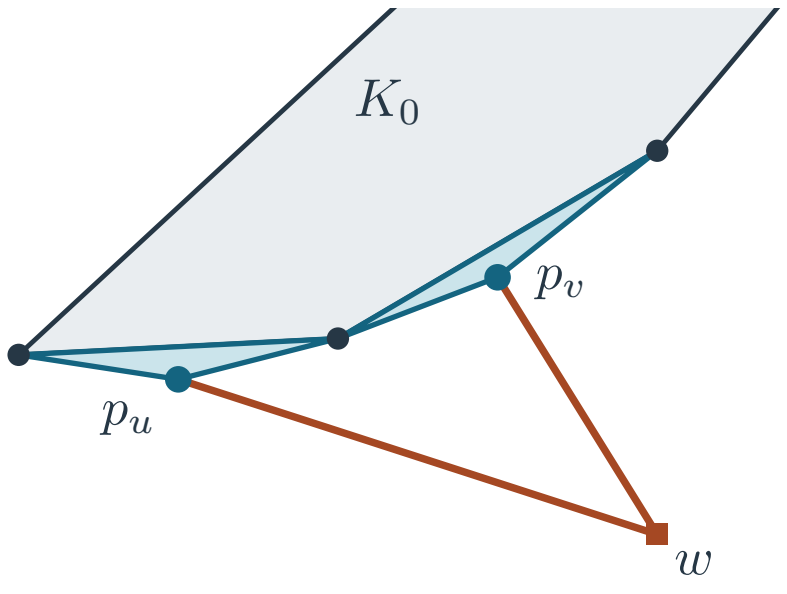

Place the common point $w=(1/2,\ldots,1/2,-1)$ below $K_0$. Figure~\ref{fig:lower-bound-overview} shows how an edge $uv$ of $G$ becomes a single input object, the chain $[p_u,w]\cup[w,p_v]$. The ray from $w$ through $p_v$ reaches $F_v$, so a convex polytope containing $K_0$ and meeting $[w,p_v]$ must contain $p_v$. The same holds for $u$. Thus every feasible polytope contains the apices of a vertex cover. Conversely, $K_0$ together with the pyramids of any vertex cover meets all input objects. The nonconvexity of the chains is essential here, since each chain encodes the disjunction ``$p_u$ or $p_v$''.

For volume, we set $h_v=h$ at every vertex, so all pyramids have the same volume. For surface area, we can choose the depths $h_v$ separately so that replacing a facet by the sides of its pyramid gives the same increase in surface area at every vertex. In either case, the common increment is $\Delta=\Theta_d(L^{-(d+1)})$, so the optimum is the cost of $K_0$ plus $\Delta$ times the size of a minimum vertex cover. Since $K_0$ has constant volume and surface area and $m=\Theta_d(L^{d-1})$, a constant gap in cover size becomes a relative gap of $\Theta_d(L^{-2})$. Taking $\varepsilon$ of this order gives $(1/\varepsilon)^{(d-1)/2}=\Theta_d(m)$, so an algorithm with the running time excluded by Theorem~\ref{thm:intro-lower} would solve the gap problem in time $2^{O(m^{1-\delta/(d-1)})}$, contradicting SH. Finally, we round the construction to rational coordinates with $O(\log n)$ bits while preserving the gap; see Section~\ref{sec:lower-bound}.
\section{Preliminaries}\label{sec:intro}

The coreset construction of Agarwal, Har-Peled, and Varadarajan~\cite{coreset} gives small subsets of finite point sets that approximate directional widths. Our structural argument needs the same kind of small representative set for a convex body, with an explicit containment guarantee. We use the following theorem of Nasz\'odi, Nazarov, and Ryabogin~\cite{NNR20}.

\begin{theorem}[\cite{coreset} and \cite{NNR20}]\label{thm:nnr}
Let $K\subset\R^d$ be a convex body with the center of mass at the origin. For every $0<\delta<1/2$, there is a polytope $P$ with at
most $ 2^{O(d)}\delta^{-(d-1)/2}$ vertices such that $(1-\delta)K\subseteq P\subseteq K.$
\end{theorem}

Our objective functions will be those computing the volume and surface area of feasible polytopes approximating the actual optimal solution. As a standard consequence of the critical-point and quantifier-elimination algorithms in \cite[Chapters~13--14]{BPR}, we have the following result. 
\begin{theorem}[Real algebraic optimization]\label{thm:bpr}
Let a semialgebraic set in $\mathbb R^p$ be given by a quantifier-free Boolean combination of $s$ polynomial predicates, each of degree at most $r$, with real coefficients. In the Real RAM model, feasibility, as well as the computation of a minimizer of a polynomial objective of degree at most $r$ whenever a minimum is attained, can be carried out using $(sr)^{O(p)}$
arithmetic operations and comparisons over the reals. 
\end{theorem}
We will use the above theorem in constant dimension. 
\paragraph{Volume and surface area.}
Let $Q$ be a full dimensional convex polytope with $m$ vertices in $\R^d$, and let
$x\mapsto Ax+t$ be an affine transformation. The volume satisfies $\Vol(AQ+t)=|\det(A)|\,\Vol(Q).$ Hence, once $\Vol(Q)$ is computed, the volume of every affine image of $Q$ can be evaluated in constant time. We compute $\Vol(Q)$ by triangulating $Q$ and summing the volumes of the resulting simplices. In fixed dimension, such a triangulation can be constructed in $O_d(m^{\lfloor d/2\rfloor})$ time; see, e.g.,
B{\"u}eler, Enge, and Fukuda~\cite{BuelerEngeFukuda00}.

For our surface area optimization problem, it is convenient to use the
equivalent facet formula
$\Surf(AQ+t)=\sum_j\|\operatorname{cof}(A)n_j\|\Vol_{d-1}(F_j)$, where $F_j$ ranges over the facets of $Q$, $n_j$ is the outward unit normal to $F_j$, and $\operatorname{cof}(A)=\det(A)A^{-T}$ is the cofactor matrix of $A$. This representation involve square roots when the Euclidean $(d-1)$-dimensional volumes are expressed algebraically. Since these square roots would lead to a number of auxiliary variables depending on the size of $Q$, we instead approximate the square root terms by a polynomial. We therefore use the facet representation, which allows us to express the resulting approximate surface area functional as a polynomial in the entries of $A$. To obtain the facet representation, we use Chazelle's algorithm~\cite{Chazelle93} to compute the half-space description $Q=\{y\in\mathbb R^d:\mathsf M y\le \mathsf h\}$, where the rows of $\mathsf M$ are the outward unit facet normals $n_j$, and the entries of $\mathsf h$ specify the corresponding supporting hyperplanes. This can be done in $O\!\left(m\log m+m^{\lfloor d/2\rfloor}\right)$ time. By McMullen's Upper Bound Theorem~\cite{McMullen70}, $Q$ has at most $O_d\!\left(m^{\lfloor d/2\rfloor}\right)$ facets, and hence the number of terms in the surface area functional is polynomial in $m$ for fixed $d$.

\section{A Grid-Based Structural Theorem}
\label{sec:structural}
In this section, we construct a grid $G$ of size $O(\eps^{-d})$ for a given $0<\eps<1$, and show that there exists a polytope with vertices in $G$ that, after applying a suitable affine transformation, is feasible and has volume $(1+\eps)\OPT_{\Vol}(\cS)$, or, respectively, surface area $(1+\eps)\OPT_{\Surf}(\cS)$.
\begin{lemma}\label{lem:rounding}
Let $0<\delta<1/2$ and $K$ be a convex body containing the unit ball such that $(1-\delta)K\subseteq P\subseteq K$ for a polytope $P=\conv\{p_1,\ldots,p\}$. 
For every $j$, let $q_j$ be a point with
$\|p_j-q_j\|\le\delta$, and $Q:=\conv\{q_1,\ldots,q_k\}.$
Then, $(1-2\delta)\B\subseteq Q\subseteq(1+\delta)K,$
and $K\subseteq(1-2\delta)^{-1}Q. $
\end{lemma}
\begin{proof}
    Observe that $P\subseteq Q+\delta\B$ and $Q\subseteq P+\delta\B$. To show $(1-2\delta)\B\subseteq Q$, take $(1-2\delta)y\in(1-2\delta)\B$ where $y\in\B$. If we have $(1-\delta)y\in Q$ then $(1-2\delta)y\in Q$. Otherwise, if $(1-\delta)y\notin Q$, consider $q\in bd(Q)$ in the direction of $y$, since $P$ contains $(1-\delta)\B$ and $P\subseteq Q+\delta\B$, we have $\|(1-\delta)y-q\|\leq\delta$, hence, $\|q\|\geq1-2\delta$, therefore, $(1-2\delta)\in Q$. On the other hand, $Q\subseteq P+\delta\B\subseteq K+\delta\B\subseteq(1+\delta)K$. Finally, 
    $$ K\subseteq\frac{1}{1-\delta}P
        \subseteq\frac{1}{1-\delta}(Q+\delta\B)\subseteq
        \frac{1}{1-\delta}
        \left(Q+\frac{\delta}{1-2\delta}Q\right)
        =\frac{1}{1-2\delta}Q.$$
\end{proof}
Observe that, as commented in \cite{NNR20}, for an arbitrary convex body $K$, there is an invertible linear transformation $T$ such that $\B\subseteq T(K)\subseteq d\sqrt{d}\,\B.$ The following theorem only uses the property that every element in our family $\cS$ is closed.
\begin{theorem}[Structural theorem]\label{thm:structural}
     For any family of closed sets $\cS$ in $\R^d$ and $\eps\in(0,1)$, there exists a polytope $Q$, with vertices on a grid $G$ of size $O(\eps^{-d})$, such that for some affine transformation $x\mapsto Ax+t$, the polytope $AQ+t$ intersects every object of $\cS$ and has volume $(1+\eps)\OPT_{\Vol}(\cS)$; the analogous statement holds for surface area, with $(1+\eps)\OPT_{\Surf}(\cS)$.
\end{theorem}
\begin{proof}
    Let $G$ be the grid obtained from the lattice $\delta\mathbb{Z}^d$ restricted to the box $[-d\sqrt{d}-\delta,d\sqrt{d}+\delta]^d$, where $\delta:=0.01\eps d^{-1}$.
    If $\OPT_{\Vol}(\cS)>0$, consider a minimum-volume intersecting polytope $K$ and assume, without loss of generality, that its center of mass is at the origin. Let $T$ be an invertible linear map such that $\B\subseteq T(K)\subseteq d\sqrt{d}\ \B$. Apply Theorem~\ref{thm:nnr} to obtain a polytope $P$ inscribed in $T(K)$ with the given properties. Round the vertices of $P$ to the grid to obtain $Q$. By Lemma~\ref{lem:rounding}, $K\subseteq T^{-1}((1-2\delta)^{-1}Q)\subseteq (1+\delta)(1-2\delta)^{-1}K$. Defining $Ax+t:=T^{-1}((1-2\delta)^{-1}x)$ we get that $AQ+t$ intersects every input object and
    $\Vol(AQ+t)\leq\left(1-2\delta\right)^{-d}\Vol((1+\delta)K)\le(1+\eps)\Vol(K)$. 

    On the other hand, if $\OPT_{\Vol}(\cS)=0$,  we set $Q$ as the rounding polytope of $[-1,1]^d$ on the grid. Consider a hyperplane restricted to a sufficiently large ball such that it intersects every set in $\cS$. This compact convex set is contained in an affine image $AQ+t$ with $\rank(A)\le d-1$, and this affine image intersects every $S_i$. 
    

  The same argument works for surface area. Namely, if  $\OPT_{\Surf}(\cS)>0$ define $K$ as a minimum-surface area intersecting polytope. Since $K$ and $Q$ are convex, $ \Surf\!\left(AQ+t\right)\le\left(1-2\delta\right)^{1-d}\Surf((1+\delta)K)\le\left(1+\eps\right)\Surf(K)$. In the case $\OPT_{\Surf}(\cS)=0$, we obtain a matrix $A$ of rank at most $d-2$ such that $AQ+t$ is feasible and has zero surface area.
\end{proof}
From now on we refer to the grid $G$ as the grid used in the previous proof.
\section{Finding an Affine Transform via Semialgebraic Optimization}\label{sec:optimization}
Given a polytope $Q=\{y\in\mathbb R^d:\mathsf M y\le \mathsf h\}$ described by $O(m^{\lfloor d/2 \rfloor})$ linear inequalities, whose vertices belong to the grid $G$, our goal is to minimize both the volume and the surface area of $AQ+t$ subject to the constraint $(AQ+t)\cap S_i\ne\emptyset$ for every $i$. We formulate this as real algebraic optimization over a semialgebraic set. The conversion takes $(s_i+m^{\lfloor d/2 \rfloor})^{O_{d,r}(1)}$ time as the following lemma also shows.
\begin{lemma}\label{lem:elimination}
The condition $(AQ+t)\cap S_i\ne\emptyset$, where $S_i\in\cS$, is equivalent to a quantifier-free Boolean formula $\Phi_{i,Q}(A,t)$ in the $d^2+d$ entries of $(A,t)$. Moreover, the formula $\Phi_{i,Q}$ contains $(s_i+m^{\lfloor d/2 \rfloor})^{O_{d,r}(1)}$ polynomial predicates, each of degree bounded by a function of $d$ and $r$. 
\end{lemma}
\begin{proof}
Let $S_i\subset\mathbb R^d$ be a semialgebraic set described by a quantifier-free Boolean formula $\Psi_i(x)$ using $s_i$ polynomial predicates of degree at most $r$. The condition $(AQ+t)\cap S_i\neq\emptyset$ is equivalent to $\exists y\in\mathbb R^d:\mathsf M y\le \mathsf h\land\Psi_i(Ay+t)$.
This formula has $d$ quantified variables $y$ and $d^2+d$ free variables
$(A,t)$. The number of polynomial predicates is $s_i+O(m^{\lfloor d/2\rfloor})$, up to a
constant factor. After substituting $x=Ay+t$ into the defining
polynomials of $S_i$, the degrees remain bounded by a function of $r$. Since $d$ is fixed, the number of quantified and free variables is bounded by a constant depending only on $d$. By effective quantifier elimination over the reals, the existential formula can be transformed into an equivalent quantifier-free formula in the variables $(A,t)$, with polynomial complexity in the number of input polynomials when $d$ and the polynomial degrees are fixed \cite[Algorithm 14.5 and Theorem 14.16]{BPR}.
\end{proof}
For a fixed grid polytope, we determine whether it has a feasible affine image and, if so, compute one that minimizes the relevant objective. Take $\delta:=0.01\eps d^{-1}$ as in the proof of Theorem \ref{thm:structural} and let $Q$ be a full-dimensional polytope as before. Define $a_j=\Vol_{d-1}(F_j)$, where $j$ runs through the number of facets $F_j$ of $Q$ and let $n_j$ be the outward unit normal to $F_j$. Given an affine transformation  $x\mapsto Ax+t$, we use the formulation $\Surf(AQ+t)=s\sum_ja_j\|Bn_j\|,$ where $B$ is a matrix of unit norm and $s:=\|\cof(A)\|$. Let $g(Q):=\sum_ja_j\|Bn_j\|$, then we can approximate $\Surf(AQ+t)$ by a polynomial that can be constructed in time $\poly(m^{\lfloor d/2 \rfloor},1/\delta,1/\eps)$, for fixed $d$.
\begin{lemma}\label{lem:polynomial-surface}
For every $0<\eps<1$, there is a polynomial $p(Q)$
of degree $\poly(m^{\lfloor d/2 \rfloor},1/\delta,1/\eps)$ such that
\[
  (1-\eps)g(Q)\leq p(Q)
  \leq(1+\eps)g(Q).
\]
\end{lemma}

\begin{proof}
Since $Q$ is full-dimensional, its facet normals span $\mathbb R^d$.
Choose $d$ linearly independent facet normals. Because the vertices
lie on the grid $G=\delta\mathbb{Z}^d\cap[-d\sqrt{d}-\delta,d\sqrt{d}+\delta]^d$, their corresponding unnormalized integer normal vectors have norms
$O_d(\delta^{-(d-1)})$. Their determinant is a nonzero integer.
Consequently, the matrix formed by the selected unit normals has
smallest singular value at least $c_d \delta^{d(d-1)}$. Moreover, each facet contains a $(d-1)$-simplex with grid vertices,
so $a_j\geq c_d \delta^{d-1}$. Hence, for every $\|B\|_F=1$, we get $g(Q)\geq c_d \delta^{d^2-1}$.
Also, since $Q$ lies in a bounded box, $\sum_j a_j=O_d(m^{\lfloor d/2 \rfloor})$.  Now, let $q$ be a polynomial approximating $\sqrt{x}$ uniformly
on $[0,1]$ with error at most $\eta$. Such polynomial exist
with degree $O(\eta^{-2})$, by the standard polynomial
approximation of the square-root function. Since
$\|Bn_j\|^2\leq1$ whenever $\|B\|_F=1$, the polynomial
\[
  p(Q)
  :=\sum_ja_jq(\|Bn_j\|^2)
\]
satisfies $|p(Q)-g(Q)|  \leq \eta\sum_j a_j.$
Taking $\eta=c_d\eps \delta^{d^2-1}m^{-\lfloor d/2\rfloor}$ with a sufficiently small
constant gives $  |p(Q)-g(Q)|\leq\eps g(Q)$.
\end{proof}
\begin{lemma}\label{lem:best-copy}
Given $Q$ and $\OPT_{\Vol}(\cS)>0$, respectively $\OPT_{\Surf}(\cS)>0$, one can either report that no affine image of $Q$ is feasible, or compute a minimum-volume feasible affine image, respectively a $(1+\eps)$-approximate minimum surface area feasible affine image.
\end{lemma}

\begin{proof}
For each object $S_i$, Lemma~\ref{lem:elimination} gives a
quantifier-free formula $\Phi_{i,Q}(A,t)$ describing feasibility.
Their conjunction has size
$(n+m^{\lfloor d/2\rfloor})^{O_{d,r}(1)}$. For volume, the identity
$\Vol(AQ+t)=|\det(A)|\Vol(Q)$ gives a polynomial objective,
so Theorem~\ref{thm:bpr} applies directly.

For surface area, introduce a scalar $s>0$ and a matrix $B$
satisfying $\operatorname{cof}(A)=sB, \|B\|_F^2=1.$
Since $\OPT_{\Surf}(\cS)>0$, every feasible affine image has
nonzero cofactor matrix, so this normalization is valid. The
surface area becomes $\Surf(AQ+t)=sg(Q)$.
By Lemma~\ref{lem:polynomial-surface}, choose a polynomial
$p(Q)$ approximating $g(Q)$ within relative error $\eta$.
We then optimize the polynomial objective $s\,p(Q)$,
subject to the feasibility constraints and the normalization
equations above. The number of variables is
$2d^2+d+1$, independent of the number of facets, and the
objective degree is polynomial in the input size and
$\eps^{-1}$. Theorem~\ref{thm:bpr} therefore gives the claimed
running time.

Choosing $\eta$ sufficiently small in terms of $\eps$ and minimizing the resulting polynomial objective to the corresponding relative precision gives a feasible image with surface area at most $(1+\eps)\OPT_{\Surf,Q}(\cS)$.
\end{proof}

Note that in the above proof, for fixed $d$ and $r$, both problems can be solved in time polynomial in $n$, the number of facets of $Q$, and
$\eps^{-1}$. In particular, both have polynomial running time
$(n+m^{\lfloor d/2\rfloor})^{O_{d,r}(1)}$ for fixed $\eps$.

Finally, we deal with the zero-volume solutions. 

\begin{lemma}\label{lem:zero}
There are $n^{O(d^2)}$-time algorithms deciding whether $\OPT_{\Vol}(\cS)=0$ or $\OPT_{\Surf}(\cS)=0$, respectively. In the affirmative cases, they return an affine hyperplane or an affine $(d-2)$-flat, respectively, meeting every set in $\cS$.
\end{lemma}
\begin{proof}
Let $Q$ be a full dimensional simplex. We have that $\OPT_{\Vol}(\cS)=0$ if and only if the following semialgebraic formula is feasible:
$$\exists(A,t)\in\R^{d^2+d}:\quad \bigwedge_i \Phi_{i,Q}(A,t)\ \wedge\ \det(A)=0.$$
Applying Theorem~\ref{thm:bpr} feasibility can be decided in time $(\sum_i s_i+m^{\lfloor d/2\rfloor})^{O(d^2}\leq n^{O(d^2)}$ time. When feasible, a sample solution $(A,t)$ is returned. Since $\det(A)=0$, the image $AQ+t$ is contained in some affine hyperplane.
Similarly, $\Surf(AQ+t)=0$ if and only if $\operatorname{rank}(A)\le d-2$, equivalently, all $(d\!-\!1)\times(d\!-\!1)$ minors of $A$ vanish. Therefore $\OPT_{\Surf}(\cS)=0$ if and only if
$$\exists(A,t)\in\R^{d^2+d}:\quad \bigwedge_{i} \Phi_{i,Q}(A,t)\ \wedge\ \operatorname{cof}(A)=\mathbf 0$$
is feasible. Again by Theorem~\ref{thm:bpr}, feasibility is decided in $n^{O(d^2)}$ time, and a sample $(A,t)$ can be returned. 

In both cases, extracting a basis of $\ker(A^\top)$ gives a defining system for the desired hyperplane or $(d-2)$-flat respectively.
\end{proof}
\paragraph{Algorithm.} We now prove Theorem \ref{thm:intro-upper}. Given the family $\cS$ in $\R^d$ and $\eps>0$, we first use Lemma \ref{lem:zero} to test whether there is a solution of zero volume, respectively zero surface area. If so, the lemma returns a hyperplane, respectively a $(d-2)$-flat, intersecting every object in $\cS$. Otherwise, let $G$ be the grid from Theorem \ref{thm:structural} and enumerate all $d$-dimensional convex polytopes $Q$ with vertices in $G$. For each such $Q$, we use Lemma \ref{lem:best-copy} to determine whether $Q$ has a feasible affine image and, if so, compute a minimum-volume or minimum-surface-area one. Theorem~\ref{thm:structural} guarantees that some such $Q$ yields a $(1+\eps)$-approximation to the actual optimums. Hence, returning the best solution found gives the desired approximation. The running time is $(nr)^{O(d^6)}\left(1/\varepsilon\right)^{O_d\left(\varepsilon^{-(d-1)/2}\right)}$ for both volume and surface area.
\newline
\newline
The polynomial exponent of $(nr)^{O(d^6)}$ is a loose bound obtained by combining quantifier elimination with fixed-dimensional real algebraic optimization~\cite[Theorem~14.16 and Algorithm~14.9]{BPR}. For each input set, the intersection test has $d$ quantified variables and $d^2+d$ free variables. The final optimization uses at most $2d^2+d+1$ variables. The exponent of $n$ is independent of $\varepsilon$, so the algorithms are efficient polynomial-time approximation schemes (EPTASes). In particular, they extend the planar schemes for convex polygons to intersecting and nonconvex semialgebraic sets, as well as to higher dimensions.

\section{Lower bound: Packing on Spheres}
\label{sec:lower-bound}

We prove lower bounds on the dependence on $\varepsilon$ in the running time of approximation schemes for \textsc{Minimum Volume Intersecting Polytope} and \textsc{Minimum Surface-Area Intersecting Polytope}. These bounds are conditional on the Satisfiability Hypothesis (SH) of Stearns and Hunt~\cite{StearnsHunt1990}. SH asserts that, for every fixed $0<\delta<1$, no algorithm solves CNF satisfiability on instances of size $s$ in time $2^{O(s^{1-\delta})}$; see also~\cite{SHhardness}. We restate Theorem~\ref{thm:intro-lower}.

\conditionalLowerBound*

Our source problem is a gap version of \textsc{Vertex Cover}. Its hardness follows from the inapproximability of Max-3SAT under SH~\cite[Corollary~5.1]{SHhardness}.

\begin{lemma}\label{lem:gap-vertex-cover}
There are constants $0<\alpha<\beta<1$ such that, assuming SH, for every
fixed $0<\delta<1$ no deterministic $2^{O(m^{1-\delta})}$-time algorithm
distinguishes an $m$-vertex graph with a vertex cover of size at most
$\alpha m$ from one in which every vertex cover has size at least
$\beta m$.
\end{lemma}

\begin{proof}
By Corollary~5.1 of~\cite{SHhardness}, we can fix a rational
$0<\rho<1$ such that SH rules out distinguishing a satisfiable
3-CNF formula with $s$ clauses from one in which every assignment
violates at least $\rho s$ clauses in time $2^{O(s^{1-\delta})}$
for any fixed $0<\delta<1$. Indeed, the constant in the exponent is
absorbed by applying the corollary with $\delta/2$. Repeat literals
as needed so that each clause has three occurrences.

Create one graph vertex for each occurrence. Join the three vertices of
each clause pairwise. Also join any two occurrences whose literals are
complementary. An independent set contains at most one occurrence from
each clause and contains no contradictory literals. Hence its literals
can be made true by one assignment. Conversely, an assignment gives an
independent set by choosing one true occurrence from every satisfied
clause. Thus the maximum independent-set size equals the maximum number
of simultaneously satisfiable clauses.

The graph has $m=3s$ vertices. Its minimum vertex-cover size is $2s$ in
the first case and at least $(2+\rho)s$ in the second. The lemma follows
with $\alpha=2/3$ and $\beta=(2+\rho)/3$. The graph has at most $O(s^2)$
edges and can be constructed in polynomial time.
\end{proof}

We use one geometric reduction for both objectives. It extends the planar construction of Dumitrescu and Jiang~\cite{DumitrescuJiang2012}.

\begin{proof}[Proof of Theorem~\ref{thm:intro-lower}]
Fix $d\ge2$ and $0<\delta<1$. Suppose that for one of the two objectives
a deterministic algorithm returns a feasible $(1+\varepsilon)$-approximation
for every $0<\varepsilon<1$ in time
\[
 T(n,\varepsilon)\le 2^{O(1/\varepsilon^{(d-1-\delta)/2})}n^{O(1)}.
\]
We use it to distinguish the two cases of Lemma~\ref{lem:gap-vertex-cover}.
Let $G=(V,E)$ be an $m$-vertex graph. Choose the even subdivision count
$L=2\lceil m^{1/(d-1)}\rceil$ and divide each coordinate of
$[0,1]^{d-1}$ into $L$ equal intervals. The resulting cells have side
length $1/L$, and $(L/2)^{d-1}\geq m$ of them lie in the first half of
the cube. Assign a different such cell to each vertex of $G$. We write
points of $\mathbb R^d$ as $(x,y)$, where $x\in\mathbb R^{d-1}$.

\paragraph{The base polytope.}
Use the convex quarter-circle arc $\gamma(t)=1-\sqrt{1-t^2}$ for
$0\leq t\leq1$. Let $q:[0,1]\to\mathbb R$ be the piecewise-linear
function with $q(j/L)=\gamma(j/L)$ for $j=0,\ldots,L$. Put
$\phi(x)=\sum_{i=1}^{d-1}q(x_i)$.
For every $x\in\{0,1/L,\ldots,1\}^{d-1}$, include
$(x,\phi(x))$ as a singleton object, and let $K_0$ be their convex hull.
Since $\gamma$ is convex, so are $q$ and $\phi$. The function $\phi$ is
affine on each cell. Thus its graph is the lower boundary of $K_0$,
with one facet above each cell. The upper boundary is the plane
$y=\sum_{i=1}^{d-1}x_i$: the cube vertices lie on this plane, and
$q(t)\leq t$ because $q$ is convex with $q(0)=0$ and $q(1)=1$.

Thus $\vol(K_0)=(d-1)\int_0^1(t-q(t))\,dt=\Theta_d(1)$. Note that
convexity and $q(1/2)=\gamma(1/2)$ give
$\int_0^{1/2}(t-q(t))\,dt\geq(\sqrt3-1)/8$, while
$\int_0^1(t-q(t))\,dt\leq1/2$.
Also $\surf(K_0)=\Theta_d(1)$: the upper facet has constant area, and
$K_0$ lies in a box of constant surface area.

For each vertex $v$, let $C_v$ be its cell, $c_v$ its center, and $F_v$
the corresponding lower facet of $K_0$. Put $h=1/(16L^2)$. Choose a
depth $h_v\in(0,h]$ and place the point
$p_v=(c_v,\phi(c_v)-h_v)$ below $F_v$. The convex hull of $F_v$ and
$p_v$ is the shallow pyramid associated with $v$.
Figure~\ref{fig:base-polytope} shows the facet above $C_v$ and the vertical depth of its apex in the planar case.

The curvature of $\gamma$ lets these pyramids remain separate. Indeed,
$\gamma''(t)=(1-t^2)^{-3/2}\geq1$ for $0\leq t<1$, so the slopes of
$q$ increase by at least $1/L$ between consecutive intervals. On the
first half of the domain, these slopes lie in
$[0,\gamma'(1/2)]=[0,1/\sqrt3]$. The next lemma uses the slope gap
to show that the pyramids contribute independently.

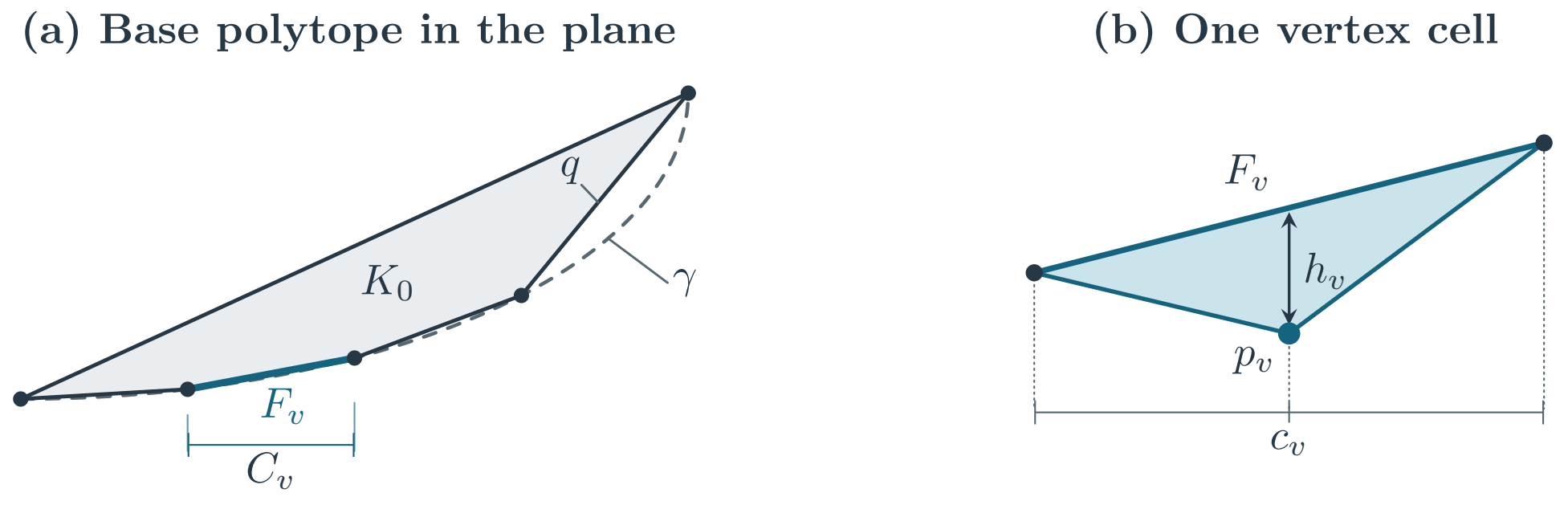
\begin{figure}[htbp]
\centering
\begin{tikzpicture}[mip picture]
  \node[mip panel] at (2.7,3.02) {(a) Base polytope in the plane};
  \node[mip panel] at (10.5,3.02) {(b) One vertex cell};
  \begin{scope}[x=5.50cm,y=2.52cm]
    \def\arcpoint#1{({#1},{1-sqrt(1-(#1)*(#1))})}
    \fill[mipBase] (0,0)--\arcpoint{.25}--\arcpoint{.5}--\arcpoint{.75}--(1,1)--cycle;
    \draw[mip boundary] (0,0)--(1,1);
    \draw[mipGray,line width=0.85pt,dashed]
      plot[domain=0:90,samples=70] ({sin(\x)},{1-cos(\x)});
    \draw[mip boundary] (0,0)--\arcpoint{.25}--\arcpoint{.5}--\arcpoint{.75}--(1,1);
    \draw[mipBlue,line width=1.65pt] \arcpoint{.25}--\arcpoint{.5};
    \foreach \t in {0,.25,.5,.75,1}
      \fill[mipInk] \arcpoint{\t} circle[radius=1.8pt];
    \node[text=mipBlue,below left=1pt] at (.45,.08) {$F_v$};
    \node at (.55,.38) {$K_0$};
    \draw[mip leader] (.84,.70)--(.865,.643);
    \node[above left=-1pt,inner sep=1pt] at (.84,.70) {$q$};
    \draw[mip leader] (.88,.525)--(.969,.38);
    \node[right,inner sep=1pt] at (.969,.38) {$\gamma$};
    \draw[mipBlue!65,thin] (.25,-.052)--(.25,-.17) (.5,-.012)--(.5,-.17);
    \draw[mipBlue,thin,|-|] (.25,-.15)--(.5,-.15);
    \node[below,inner sep=1.7pt] at (.375,-.15) {$C_v$};
  \end{scope}
  \begin{scope}[xshift=8.35cm]
    \coordinate (left) at (0,1.04);
    \coordinate (right) at (4.2,2.11);
    \coordinate (mid) at (2.1,1.575);
    \coordinate (apex) at (2.1,0.54);
    \path[mip pyramid] (left)--(apex)--(right)--cycle;
    \draw[mipBlue,line width=1.35pt] (left)--(right);
    \draw[mipGray,densely dotted] (0,0.99)--(0,-0.08)
      (4.2,2.04)--(4.2,-0.08) (2.1,0.48)--(2.1,-0.08);
    \draw[mipGray,thin,|-|] (0,-0.11)--(4.2,-0.11);
    \draw[mipGray,thin] (2.1,-0.03)--(2.1,-0.19);
    \node[below,inner sep=1.8pt] at (2.1,-0.19) {$c_v$};
    \draw[<->,>=stealth,mipInk,line width=0.65pt,shorten <=2pt,shorten >=1pt]
      (apex)--(mid) node[midway,right=1pt] {$h_v$};
    \fill[mipInk] (left) circle[radius=2pt] (right) circle[radius=2pt];
    \fill[mipBlue] (apex) circle[radius=2.6pt];
    \node[above left=2pt] at (mid) {$F_v$};
    \node[below left=1pt] at (apex) {$p_v$};
  \end{scope}
\end{tikzpicture}%
%
\caption{The base polytope and one vertex pyramid in dimension two. (a)~Joining consecutive samples of the dashed arc $\gamma$ gives the lower boundary $q$ of $K_0$. The highlighted segment lies above the cell $C_v$. (b)~Above its midpoint $c_v$, the apex $p_v$ lies a vertical distance $h_v$ below the facet $F_v$. Selecting $v$ adds the shaded triangle. The depth is enlarged for visibility.}
\label{fig:base-polytope}
\end{figure}

\begin{lemma}\label{lem:independent-pyramids}
For any choice of the depths $h_v\in(0,h]$ and any $U\subseteq V$, let
$K_U$ be the convex hull of $K_0$ and $\{p_v:v\in U\}$. Then $K_U$ is the
union of $K_0$ and the pyramids with bases $F_v$ and apices $p_v$, for
$v\in U$. In particular,
$\vol(K_U)=\vol(K_0)+\sum_{v\in U}h_v/(dL^{d-1})$.
For surface area, each selected pyramid replaces its base facet by its
side facets, independently of the other selected pyramids.
\end{lemma}

\begin{proof}
Give depth zero to every cell that does not correspond to a vertex in
$U$. On a cell $C$ with center $c$ and depth $h_C$, set
$f(x)=\phi(x)-h_C+2Lh_C\|x-c\|_\infty$. This is the lower boundary of
the proposed union. It is convex within each cell and agrees with
$\phi$ on the cell boundary.

Consider two cells sharing a face. The normal slope of $\phi$ increases
by at least $1/L$ across this face. The two depth terms can decrease this
increase by at most $4Lh=1/(4L)$. The tangential slopes agree because
both pieces equal $\phi$ on the face. Thus $f$ is convex on the whole
cube. The region between its graph and the upper plane of $K_0$ is
convex. It is precisely the stated union, so it is also $K_U$.

Each pyramid has volume $h_v/(dL^{d-1})$. Its interior is disjoint from
$K_0$ and from the other pyramids. Its side facets replace $F_v$ on the
boundary of $K_U$. This proves both assertions.
\end{proof}

\paragraph{The edge objects.}
Let $w=(1/2,\ldots,1/2,-1)$. For each edge $uv\in E$, add the chain
$[p_u,w]\cup[w,p_v]$. The next lemma shows that meeting either segment
forces the corresponding apex. Figure~\ref{fig:forced-endpoint} shows how convexity forces $p_v$ on the arm $[w,p_v]$.

\begin{figure}[htbp]
\centering
\begin{tikzpicture}[mip picture,x=1.22cm,y=0.96cm]
  \coordinate (left) at (1.2,1.50);
  \coordinate (right) at (5.2,2.20);
  \coordinate (w) at (7.52,-0.90);
  \path (left)--(right) coordinate[pos=0.23] (bv);
  \path (bv)--(w) coordinate[pos=0.20] (pv) coordinate[pos=0.62] (t);
  \begin{scope}
    \clip (0,1.35) rectangle (8.1,3.40);
    \path[fill=mipBase,draw=mipInk,line width=0.8pt] (0.10,1.48)--(left)--(right)--
      (8,3.72)--(14.94,10)--cycle;
  \end{scope}
  \path[mip pyramid] (left)--(pv)--(right)--cycle;
  \node at (3.45,2.53) {$K_0$};
  \node[above=2pt,text=mipBlue] at (4.27,2.04) {$F_v$};
  \draw[mipInk,dashed,line width=0.9pt] (bv)--(pv);
  \draw[mip chain] (pv)--(w);
  \fill[mipInk] (bv) circle[radius=2.4pt];
  \fill[mipBlue] (pv) circle[radius=2.7pt];
  \node[circle,draw=mipRust,fill=white,line width=1.0pt,inner sep=2.0pt] at (t) {};
  \node[mip bend] at (w) {};
  \node[above left=1pt] at (bv) {$b_v$};
  \node[below left=1pt] at (pv) {$p_v$};
  \node[above right=1pt] at (t) {$t$};
  \node[right=3pt] at (w) {$w$};
\end{tikzpicture}%
%
\caption{Why meeting an edge segment forces its apex. In this local view, the dashed extension of $[w,p_v]$ meets the facet $F_v$ at $b_v$. If a convex set $K$ containing $K_0$ meets the segment at $t$, it contains both $t$ and $b_v$. The apex $p_v$ lies between them and hence belongs to $K$.}
\label{fig:forced-endpoint}
\end{figure}
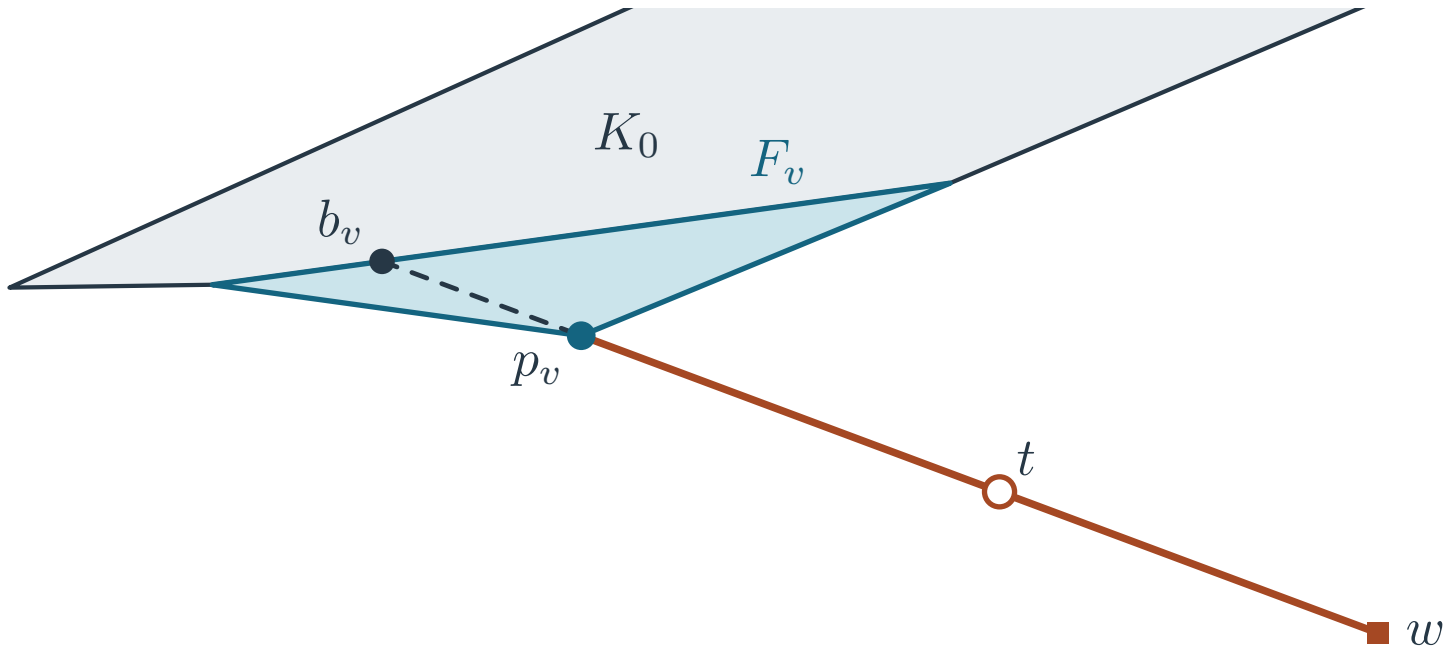

\begin{lemma}\label{lem:forced-endpoint}
Let $K$ be a convex set containing $K_0$. If $K$ meets $[p_v,w]$, then
$p_v\in K$.
\end{lemma}

\begin{proof}
Extend the plane of $F_v$ beyond its cell. At the horizontal position
of $w$, this plane has nonnegative height. Indeed, that position lies to
the right of $C_v$ in every coordinate, and all slopes of $\phi$ are
nonnegative. The vertical distance of $p_v$ below the plane is $h_v$;
the point $w$ lies at least one unit below it.

Extend the ray from $w$ through $p_v$ until it reaches the plane. The
extension beyond $p_v$ is less than $h_v/(1-h_v)<2h$ times the length of
$[w,p_v]$. The horizontal coordinates of $w$ and $p_v$ differ by at most
$1/2$, so the intersection point is less than $h$ from $c_v$ in each
coordinate. Since $h<1/(2L)$, the ray meets $F_v$ at a point $b_v$ above the interior of $C_v$.
For any $t\in K\cap[p_v,w]$, the apex $p_v$ lies on $[b_v,t]$.
Since $b_v,t\in K$ and $K$ is convex, $p_v\in K$.
\end{proof}

\paragraph{The cost of one vertex.}
For volume, take $h_v=h$ for every vertex. By
Lemma~\ref{lem:independent-pyramids}, each selected vertex increases the
volume by the same amount $h/(dL^{d-1})$.

For surface area, we choose the depths so that the increments are equal.
For $0\leq t\leq h$, let $D_v(t)$ be the surface-area change when the
apex below $F_v$ is placed at depth $t$. This function is continuous and
$D_v(0)=0$.

\begin{lemma}\label{lem:surface-increment}
For every assigned cell and $0<t\leq h$,
$D_v(t)=\Theta_d(L^{-(d-1)}(Lt)^2)$, with constants independent of the cell and of $L$.
Moreover, $D_v$ is strictly increasing. In particular,
$D_v(h)=\Theta_d(L^{-(d+1)})$.
\end{lemma}
\begin{proof}

Let $s$ be the slope vector of $F_v$ and put
$g(z)=\sqrt{1+\|z\|^2}$. Each of the $2(d-1)$ side facets projects onto
a region of $(d-1)$-volume $1/(2(d-1)L^{d-1})$. Let $e_i$ be the
$i$th standard unit vector of $\mathbb R^{d-1}$. The slopes of the
opposite facets in direction
$i$ are $s+2Lt e_i$ and $s-2Lt e_i$.
Subtracting the area $L^{-(d-1)}g(s)$ of the base gives
\[
 D_v(t)=\frac{1}{2(d-1)L^{d-1}}
 \sum_{i=1}^{d-1}\bigl(g(s+2Lt e_i)+g(s-2Lt e_i)-2g(s)\bigr).
\]
Here $s\in[0,1/\sqrt3]^{d-1}$, and
$\frac{d^2}{dz^2}g(s+ze_i)
=(1+\sum_{j\ne i}s_j^2)/(1+\|s+ze_i\|^2)^{3/2}$.
For $|z|\leq2Lh=1/(8L)$, we have $|s_i+z|<1$ and
$\|s+ze_i\|^2\leq d-1$. The numerator of the derivative is between
$1$ and $d$. Thus the derivative lies between $d^{-3/2}$ and $d$.

Each paired
difference in the displayed sum is consequently between
$d^{-3/2}(2Lt)^2$ and $d(2Lt)^2$. It follows that
$D_v(t)=\Theta_d(L^{-(d-1)}(Lt)^2)$ for $0<t\leq h$, and in particular
$D_v(h)=\Theta_d(L^{-(d+1)})$. Positivity of the same second derivative
shows that $D_v$ is strictly increasing.
\end{proof}

Choose the common surface-area increment to be
$\tfrac12\min_{v\in V}D_v(h)$. By continuity, each vertex has a depth
$h_v\in(0,h)$ at which its surface-area increase equals this value.
Write $\Delta$ for the common per-vertex increment of the chosen
objective. In either case,
$\Delta=\Theta_d(L^{-(d+1)})$.

\paragraph{Correctness of the reduction.}
Let $A$ be the volume or surface area of $K_0$, according to the chosen
objective. If $U$ is a vertex cover, then $K_U$ meets every edge chain:
for each edge $uv$, it contains $p_u$ or $p_v$. It also contains all
singleton objects. Lemma~\ref{lem:independent-pyramids} gives its cost
as $A+|U|\Delta$.

Conversely, let $K$ be any feasible convex polytope. The singleton
objects imply $K_0\subseteq K$. Put $U=\{v:p_v\in K\}$. Since $K$ meets
every edge chain, Lemma~\ref{lem:forced-endpoint} implies that $U$ is a
vertex cover. Moreover, $K_U\subseteq K$. Volume and surface area are
monotone under inclusion for convex bodies. Hence the cost of $K$ is at
least $A+|U|\Delta$. Let $\tau(G)$ denote the minimum vertex-cover
size. We conclude that the optimum is $A+\tau(G)\Delta$.

\paragraph{Rational coordinates.}
The construction above uses values of the quarter-circle arc that may
be irrational. For the actual reduction, round each interior value
$\gamma(j/L)$ to the nearest multiple of $L^{-8}$, keeping $\gamma(0)=0$
and $\gamma(1)=1$ exact. Each slope of the resulting interpolant changes
by at most $2L^{-7}$, and each gap between consecutive slopes changes by
at most $4L^{-7}$. The gaps therefore remain at least $1/(2L)$.
This is enough for the independent-pyramid argument, since the depth
terms change a slope gap by at most $4Lh=1/(4L)$. The slopes on the
assigned cells remain in $[0,1]$, and the bounds for the base and the
surface-area increments remain valid. From here on, $A$ and $\Delta$ refer to the rounded base.

The rounded samples can be found by integer square-root comparisons.
For volume the depth $h$ is already rational. For surface area, recall
that $\Delta=\tfrac12\min_v D_v(h)$. Each comparison of two values in
the formula for $D_v$ involves only a constant number of square roots
in fixed dimension, so it takes polynomial bit time. We can therefore
find the minimum and bisect $[0,h]$ for each $v$, comparing $D_v(t)$
with $\Delta$. Since $g(z)=\sqrt{1+\|z\|^2}$ is $1$-Lipschitz, the
formula gives $|D_v(t)-D_v(t')|\le 2L^{-(d-2)}|t-t'|$.
After $O_d(\log L)$ bisections the upper endpoint is a positive rational
depth with $|D_v(h_v)-\Delta|\le L^{-(d+5)}$. All coordinates thus
have $O_d(\log m)$ bits. By Lemma~\ref{lem:independent-pyramids}, for
every $U\subseteq V$ the cost of $K_U$ differs from $A+|U|\Delta$ by
at most $R$, where $R=0$ for volume and
$R=mL^{-(d+5)}=O_d(L^{-6})$ for surface area.

\paragraph{Choosing the approximation parameter.}
The two cover thresholds give costs $B=A+\alpha m\Delta$ and
$B+\Gamma$, where $\Gamma=(\beta-\alpha)m\Delta$. Since $m=\Theta_d(L^{d-1})$ and
$\Delta=\Theta_d(L^{-(d+1)})$, we have $B=\Theta_d(1)$ and
$\Gamma=\Theta_d(L^{-2})$. Fix a sufficiently small positive dyadic
constant $c_d$ and set $\varepsilon=c_d/L^2$. For all sufficiently
large $m$, this choice gives $0<\varepsilon<1$,
$\varepsilon B\le \Gamma/4$, and $R\le \Gamma/16$. The parameter has
$O_d(\log m)$ bits. The finitely many smaller graphs can be solved
by exhaustive search.

Run the assumed algorithm and let $K$ be its output. If $G$ has a
cover $U$ of size at most $\alpha m$, the feasible polytope $K_U$
costs at most $B+R$. Hence
\[
  \operatorname{cost}(K)\le(1+\varepsilon)(B+R)
     \le B+\Gamma/4+2R < B+\Gamma-R .
\]
Let $U_K=\{v:p_v\in K\}$. Since $K_{U_K}\subseteq K$, any output
with $|U_K|\ge\beta m$ costs at least $A+\beta m\Delta-R=B+\Gamma-R$.
Thus $|U_K|<\beta m$ in the first case. In the second case
$|U_K|\ge\beta m$ because $U_K$ is a vertex cover. Counting the input
apices in $K$ therefore distinguishes the two cases.

\paragraph{Running time.}
The instance has $(L+1)^{d-1}=O_d(m)$ singleton objects and at most
$\binom m2$ two-segment chains. Each has a description with $O_d(1)$
linear predicates whose coefficients have $O_d(\log m)$ bits.
The entire instance has $O_d(m^2\log m)$ bits and is constructed in
deterministic polynomial time for fixed $d$. In particular,
$n=m^{O(1)}$. Put $\eta=\delta/(d-1)$.
Since $L=\Theta_d(m^{1/(d-1)})$ and $\varepsilon=c_d/L^2$,
\[
  (1/\varepsilon)^{(d-1-\delta)/2}
    =c_d^{-(d-1-\delta)/2}L^{d-1-\delta}
    =\Theta_{d,\delta}\!\left(m^{1-\delta/(d-1)}\right).
\]
Substitution into the assumed bound gives
$T(n,\varepsilon)\le 2^{O(m^{1-\eta})}m^{O(1)}$.
The $m$ point-in-polytope tests take time polynomial in $m$ and the
length of the explicit vertex or facet description of $K$ in fixed
dimension. That length is at most $T(n,\varepsilon)$. Including
construction and decoding, the deterministic running time is therefore
\[
  m^{O(1)}+\bigl(m+T(n,\varepsilon)\bigr)^{O(1)}
       =2^{O(m^{1-\eta})}.
\]
Here $0<\eta<1$, so $m^{1-\eta}$ absorbs the logarithm of every
polynomial factor. This contradicts Lemma~\ref{lem:gap-vertex-cover}
with parameter $\eta$ and proves the stated bound for each objective.
\end{proof}

\section*{Use of AI}
The authors used GPT Astra~6.2 to draw the figures, improve the exposition, and assist in filling in technical details in the arguments.

\bibliographystyle{plain}
\bibliography{references}

\end{document}